# Assessing the Effects of Illuminance and Correlated Color Temperature on Emotional Responses and Lighting Preferences Using Virtual Reality

**Armin Mostafavi [1,2], Tong Bill Xu [2], Saleh Kalantari [2]**

[1] Research Department for Digital Architecture and Planning, Vienna University of Technology, Karlsplatz 13, A-1040 Vienna, Austria.

[2] Department of Human Centered Design, Cornell University, Ithaca, NY, USA.

**Abbreviations:**

CB Cold-Bright
CD Cold-Dark
DST Digit Span Test
EEG Electroencephalography
ET Experiment Time
HDR High Dynamic Range
HMD Head-Mounted Display
LDR Low Dynamic Range
LC Lighting Condition
MEQ Morningness–Eveningness Questionnaire
NM Neutral-Moderate
RQ Research Question
TMO Tone Mapping Operator
SAM Self-Assessment Manikin
UE4 Unreal Engine 4
VE Virtual Environment
VR Virtual Reality
WB Warm-Bright
WD Warm-Dark

**Abstract:**

This paper presents a novel approach to assessing human lighting adjustment behavior and preference in diverse lighting conditions through the evaluation of emotional feedback and behavioral data using VR. Participants (n= 27) were exposed to different lighting (n=17) conditions with different levels of illuminance and correlated color temperature (CCT) with a randomized order in a virtual office environment. Results from this study significantly advanced our understanding of preferred lighting conditions in virtual reality environments, influenced by a variety of factors such as illuminance, color temperature, order of presentation, and participant demographics. Through a comprehensive analysis of user adjustment profiles, we obtained insightful data that can guide the optimization of lighting design across various settings.

**Keywords: CCT, Illuminance, Lighting, Order effects, Gender, Virtual Reality**

1. Introduction

The exploration of how individuals respond to light in a comprehensive manner has the potential to yield significant insights that could improve various aspects of human wellbeing, including circadian health, architectural design, cognitive function, and overall quality of life [1], [2]. Lighting conditions profoundly impact the effectiveness and satisfaction of activities within architectural spaces, particularly in areas where visual engagement and cognitive tasks are crucial [3], [4]. Therefore, an area that warrants further investigation is a meticulous examination of how lighting influences individuals' perceptions and behaviors in the context of work-related activities, such as within an office environment.

Humans react to light in a variety of ways, both visually and non-visually, or image-forming (IF) and non-image-forming (NIF) pathways [5]–[10]. The processing of light information by the visual system is referred to as the IF pathway. Important pathways to people's psychological functioning and wellbeing are presented in this visual pathway. However, since the discovery of the fifth photoreceptor, known as intrinsically photoreceptive retinal ganglion cells (ipRGCs), the NIF pathway has received a lot of attention. The ipGRCs appear to be the primary cause of the effects of light on our internal clock and alertness-related areas in the brainstem and central nervous system [11], [12] as well as physiological responses. Human-centric lighting (HCL) takes into account both the IF and NIF impacts of light to promote beneficial human outcomes such as visual, circadian, and neurobehavioral [13]. It should be minded that it is challenging to implement a successful human-centric lighting solution. Therefore, this work aims to investigate and better understand individuals' nuanced behavioral and emotional responses to various lighting conditions by providing insights into how different lighting scenarios influence human interaction and emotional states within a controlled virtual environment.

## 1.1. Correlated Color Temperature and Illuminance Correlation

In terms of human mood effects, correlated color temperature (CCT) and illuminance level are the two most crucial components of light to take into account [14], [15]. Kruithof curve, which was introduced in 1941, depicts a range of illuminance levels and color temperatures that observers find appealing, agreeable, or desirable [16]. Despite the failure of multiple attempts to validate the curve, the Kruithof graph is arguably the most often reproduced graph in lighting history and has been a part of lighting guidelines for many years [17], [18]. During past decades, the preferred combination of illuminance and CCT threshold as well as the effect dominancy of these two variables on the Kruithof curve, are investigated. A revised Kruithof graph as suggested by Fotios [19], concludes that CCT changes have little impact on brightness and pleasantness ratings. Kruithof's rule indicates that individuals tend to feel more comfortable and find low illuminance and low CCT lighting conditions pleasing. Similarly, high illuminance and high CCT conditions are also associated with comfort and aesthetic appeal. This rule has been widely utilized in the field of interior lighting design [20]–[23]. However, subsequent studies have yielded contradictory findings, leading to significant controversy surrounding the validity and applicability of Kruithof's rule. For instance, Davis and Ginthner [21], stated that light illuminance level (not color temperature) was the most important component influencing subjective preference ratings. Similarly, Boyce and Cuttle's research [20] found that after individuals have become used to the lighting, the lamp's color temperature has no effect on how they perceive the room's illumination. In investigating the correlation between the spatial distribution of light and subjective feelings of comfort and emotional state, some studies (e.g.,[5], [24]) have advocated for lower CCT. Conversely, other research (e.g., [25], [26]) has indicated a preference for higher CCT.

Despite the pleasantness, human performance and subjective behavior in given lighting conditions with different illuminance levels and CCTs is a crucial aspect of lighting research in a working and educational environment. For instance, in an office environment, lighting has been shown to substantially impact work performance since it is strongly related to visual acuity, mood, and alertness [27]–[29]. In the case of illuminance level, a higher light level might enhance workers' visual performance, resulting in a lower error rate, or it might result from increased alertness, which can cut accident and error rates [30], [31]. Illuminance level also affects different types of human memories, suggesting that the effect of illuminance level on cognitive task performance may be influenced by task complexity and type (i.e., short-memory, long memory, comprehension, etc.) [10], [32], [33]. Huiberts and colleagues' study [34] showed that bright (1700 lx vs. 600lx vs. 165 lx) light exposure could improve user performance on a difficult working memory task (i.e., Backwards Digit-Span Task of 7-8 digits) while it was not the case for the easy trials (4-6 digits). Thus, task type and task difficulty level appear to be moderators of the influence of illumination on cognitive performance [35].

Lighting CCT has drawn much attention to the influence of lighting on human psychological processes [36] and leads to a variety of improvements in users' wellbeing, functioning, and task performance [9] CCT of light has been shown to improve cognitive function [37], physiological and brain dynamics [38]–[40], subjective alertness, performance, and evening fatigue [14], [41], as well as wellbeing and productivity in the corporate setting [9], [42]. Researchers also found that NIF effects are expected to be stronger when exposed to monochromatic blue light or polychromatic light with a high CCT [43], [44] However, several empirical studies found no or little effects of CCT on subjective mood and performance, and these effects found in other research studies may actually be due to illumination differences [45]–[48].

**1.2 Virtual Reality and Lighting Studies**

Researchers can test lighting hypotheses within a created 3D model and get quick user input by utilizing the visual immersion offered by VR head-mounted displays (HMDs) [49]–[53]. The VR method enables instantaneous switching between several CCTs and illuminances without incurring construction costs or posing any other logistical challenges. This makes it possible to create a highly controlled environment that can also be easily connected with questionnaires that are frequently too time-consuming, distracting, or technologically impractical to utilize in real-world scenarios. Moreover, recording additional human behavior (i.e., lighting adjustment button pressed order, precise interaction timestamps, eye-tracking, etc.) can add another layer of empirical evidence to improve our understanding of how people react to lighting and compare it to more established measures of participant behavior and subjective reaction.

The use of VR in lighting research has increased in recent years, while many studies discovered similar human reactions to lighting in virtual reality environments and real settings [54]. Studies have compared lighting perception in virtual reality and real-world surroundings and suggested that VR could be useful in testing visual perception in HCL [55]–[62]. Different studies used VR for lighting evaluation in architectural spaces, such as increasing luminance area (windows) and viewing point [60], [63]–[65], CCT and illuminance perception [33], [45], [66] or automated lighting system and energy consumption [57], [67]. Despite these encouraging results, it is crucial to understand that, of yet, the range of luminance that VR HMDs can create is still constrained when compared to actual light sources. As a result, the utility of the VR approach decreases at the extremes of both very high and very low illuminance ends, which are normally restricted to peak brightness values of less than 200 nits [68], [69].

Even with the mentioned limitation for VR-lighting research, the former studies also support that human perception and judgment in VR were similar in most lighting conditions in the real environment. For instance, the findings in Chamilithori and colleagues' experiment [55] demonstrate a high degree of perceptual accuracy, with no discernible changes in the evaluations between the real and virtual settings. The individuals reported feeling very present in the virtual world, and using the VR headset had no significant impact on their physical symptoms. Similarly, in Rockcastle and colleagues' experiment [61], results showed that when analyzing well-lit (not too dark nor overly contrasted) electric lighting scenes for subjective judgments about visual comfort, pleasantness, evenness, contrast, and glare rating, VR HMDs can be a reasonable substitute for real-world lighting conditions.

**1.3 Preferred Lighting Condition and Gender**

The user preferences through illuminance adjustment, and CCT has been widely investigated [70]–[75]. An anchor point is the starting point or stimuli encountered before judgment [76], and can alter the user preference for illuminance or CCT during an adjustment task. For example, in a study by Logadottir and colleagues [70] the experiment's findings indicated that the stimulus range and anchor substantially impacted the adjustment task's outcome. The result showed that participants have a stimulus range bias; as the maximum available illuminance in a range was raised, then the appraisal of preferred illuminance also increased. Similarly, the study by de Kort and colleagues [77] showed that pre-set values with higher numbers result in higher preferred numbers. However, in this study, the perceived level of comfort after adjustment is not dependent on the pre-set value of illuminance, and participants reported a higher visual comfort after adjustment. The adaption time before the lighting adjustment task is also shown to affect the final preferred lighting illuminance [70] and CCTs [71] generally showing that non-adapted participants tend to adjust the light intensity (lower than average 500lx on the surface) and CCTs higher than adapted (i.e., after 5 minutes of exposure to the given LC). It should be noted that, in the current study, we are not interested in the preferred values of the adjusted scene to classify comfortable vs uncomfortable conditions as the adjustment task is not an appropriate method for determining optimum lighting conditions because of the variability of the given anchor points [78].

The preferred lighting conditions in work environments have also been extensively investigated through research studies [70], [72], [78], [79]. These studies aim to understand the impact of lighting on factors such as productivity, wellbeing, and visual comfort. However, a few studies have explored potential gender differences in lighting preferences [80], [81]. A study by Kim and colleagues [82] showed that women are significantly more likely to express dissatisfaction with indoor environment quality (IEQ) factors such as lighting and visual comfort. This suggests that there may be differences in the perception of lighting conditions between genders in certain contexts. Another study by Knez and colleagues [83] showed that the lighting in the room had an impact on the participants' mood. Specifically, it influenced their negative mood, but this effect varied based on age rather than gender. Among the younger participants, it was observed that they were most successful in maintaining their negative mood when exposed to warm (reddish) white lighting, while they struggled to do so under cool (bluish) white lighting during the 90-minute cognitive task session. Lee and colleagues [84] reported significant gender differences in emotional reaction and cognitive performance to interior lighting conditions. Women generally perceived atmospheric attributes, such as colors or lighting conditions, more prominently and showed more sensitive reactions compared to men. Bodrogi and colleagues [85] examined how observer preferences for perceived illumination chromaticity were influenced, and they discovered that gender had a significant impact on the preference for various correlated color temperatures. This finding suggests that gender may contribute to the formation of preferences for lighting chromaticity.

### 1.4. User Preference Profile

User control over lighting conditions has been extensively studied, and several benefits have been identified in various domains, such as personalization[57], [86]–[88], task-specific lighting [89]–[92], energy efficiency [29], [89], [93], and patient comfort [94]–[97]. Therefore, exploring the adjustment activities and user control actions during the adjustment task is a logical approach to deriving a comprehensive and accurate user preference profile. Analyzing how users interact with and modify lighting conditions makes it possible to gather valuable data that serves as a reliable indicator of their preferences. This exploration aims to understand the patterns and behaviors exhibited by users when adjusting lighting, providing insights into their individual preferences and needs. By developing a detailed user preference profile, we can better enhance our understanding of what factors contribute to user satisfaction and tailor lighting solutions to meet their requirements. For instance, Despenic and colleagues [98] introduced a method for modelling lighting preference profiles of users based on their control behavior and preference information. Their results showed significant differences between the lighting preference profiles of users based on personal characteristics. Boyce and colleagues [91] provided evidence of significant discrepancies in control behavior among occupants. Their findings indicate that individuals vary greatly in their approach to adjusting illuminance levels. While some individuals make slight adjustments within a limited range, others exhibit a more extensive range of illuminance level adjustments. However, occupants are not characterized by illogical or irrational behavior; instead, they strive to restore their comfort using the most convenient means available to them [99] This logical reasoning highlights the importance of studying user control actions to uncover and document valuable information about user preferences to inform the design and implementation of effective lighting solutions.

### 1.5 Study Goals and Research Questions

In this work, we developed and evaluated a virtual reality platform to replicate various lighting situations having different illuminance levels and CCT. Users were exposed to these settings and could modify the lighting attributes in a simulated office room scenario. The study focused on the user adjustment behavior for each setting and the potential use of virtual reality technologies in future lighting investigations.

To assess the arousal and valence of the acute impact of illuminance and CCT, user behavior, and impressions of the technology, we used a range of quantitative assessment devices (described in more detail below). In addition, we obtained qualitative data in the form of an interview to learn more about the participants' experiences. The comparison of physical versus virtual illumination modeling and validation was not addressed in this study; this will be investigated in future research. The purpose of this research was to examine three main research questions (RQs):

**RQ1:** How will Preferred Lighting Conditions (PLC) (brightness and color temperature) be affected by Initial Lighting Conditions (ILC) (RQ1a)? Based on former studies on the lighting adjustment task [71], [78], [100], [101] we decided to investigate an expanded gradation of lighting conditions in VR while considering the user adjustment interaction behavior for setting the preferred lighting illuminance level and CCT. Moreover, as former studies suggested [80], [83], [84], [102], [103], the preferred illuminance and CCT could be different by gender; we also investigated those preferences moderated by gender.

**RQ2:** What influence will the order effect of the presented lighting conditions have on the preferred lighting condition in VR? Our study examines how the sequence of light exposure impacts participants' preferences for lighting characteristics. In a design involving repeated measures[104], [105], where observers make adjustments to multiple levels of discomfort, it is natural to anticipate that the current setting may be influenced by previous ones, known as an order effect. These order effects are associated with various factors that can confound the experimental results due to the sequence of evaluations performed by the participants [104]–[107]. Therefore, we seek to determine if the preferred lighting characteristics change after experiencing different lighting conditions or adapting to a VR-HMD. Former studies [70], [71], [78], [108], [109] on the lighting adjustment task suggest that when participants were presented with a lower range, it resulted in significantly lower illuminances. These findings from the experiment confirmed that the stimulus range, anchor point, and lighting order significantly influenced the preferred CCT determined through the adjustment task. Therefore, it is crucial to consider and report this information in studies that utilize the adjustment method, highlighting its importance for accurately interpreting results.

**RQ3:** How will the self-reported level of arousal and emotional valence be affected by given illuminance and CCT of the ILC? And similar to RQ1, are these effects moderated by gender? While former research shows the profound effect of light intensity and CCT on reported emotion [110], [111], we are looking for a replication of those results in a wide range of lighting conditions in VR.

In addition to our three RQs, to leverage the VR technology, we were interested in exploring the user adjustment pattern for preferred lighting conditions, which is not feasible or cumbersome to measure in real lab environments. In other words, how does the lighting preference profile of users in the given virtual environment differ in different lighting conditions? The former research [91], [98], [112]–[114] suggests that the level of activity exhibited by each user can be assessed by analyzing the number of user control actions. User control actions serve as a reliable indicator for deriving the user's preference profile. This research question intends to develop and report the adjustment activities during the adjustment task to derive a comprehensive and accurate user preference profile.

## 2. Research Design and Methods

### 2.1 Virtual Reality Development

The VR testing set was created as a contemporary interior office space with the following measurements: 3.7 m x 7.2 m x 3.0 m (Fig. 1). Autodesk 3ds Max was used for most of this environment's modeling and UV mapping. Unreal Engine 4.26 (UE4) was used to create the final environment in VR and implement the interaction, lighting setting, and texturing. The Blueprint visual scripting was used for all user front-end interactions. From seated at a desk, participants experienced the virtual office setting and had free access to glance around and observe various parts of the space. Participants were permitted to move their heads and body freely within the environment and change the camera angle. Still, the camera height was related to the placement of the HMD in space (equivalent to the typical human eye height in a seated state), and teleportation was not permitted.

All the participant behavior and interactions were recorded in a log file which was generated using the participant ID at the start of the VR session. The timestamps (in milliseconds) for each event and all keystroke dynamics were included in the UE4-generated log file. These keystroke adjustments were also associated with the corresponding values for lumen intensity, CCT, start-end task label, and emotional responses throughout the light adjustment scenario. Participants experienced the VR environment sitting in the middle of the office with ~6.5m distance from the screen using the Meta Quest 2 head-mounted display connected to a gaming desktop computer, with each eye having a resolution of 1832 x 1920 pixels and a maximum brightness of 100 nits. Meta Quest 2 can be adjusted for users with different interpupillary distances and offers a horizontal field of view of 90° and a refresh rate of 90 Hz.

**Figure 1.** A study participant is wearing a VR head-mounted display in the virtual office.

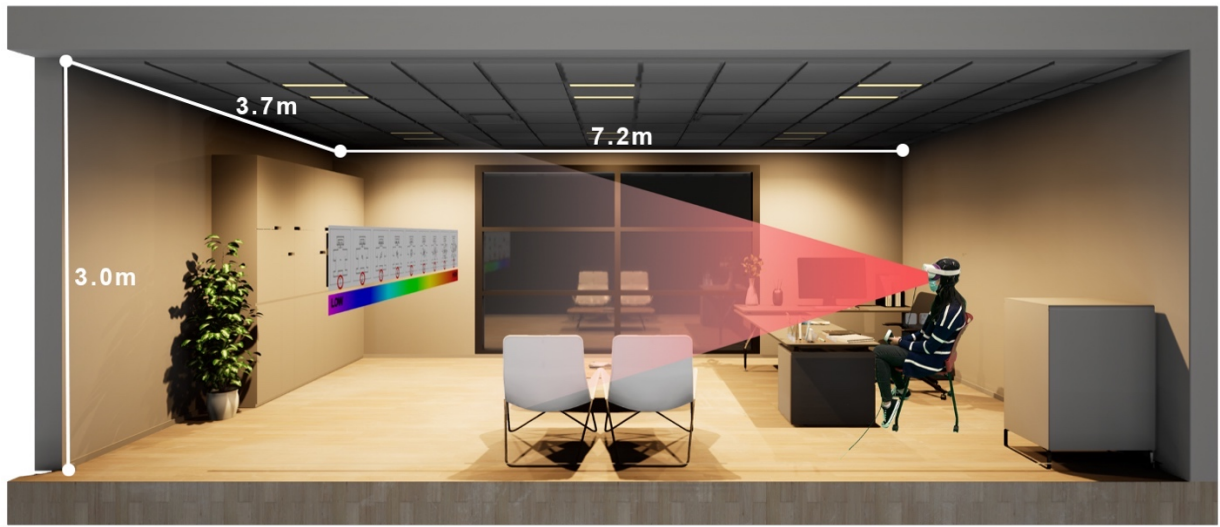

## 2.2 Lighting Conditions

The pre-processing lighting simulation was conducted using DIALux Evo software with the CoreLine Recessed profile. Seventeen lighting conditions were created using six luminaires ranging from 1500 K to 7500 K in CCT and 300lm to 4500lm in luminous flux. The corresponding illuminance for each lighting condition measured on 1m working plane is reported in Table 1. The creation criteria was based on having a series of lighting conditions with a) constant illumination with different CCTs (7 conditions), b) constant CCT with different illumination intensities (7 conditions), and c) "quadrant" conditions consisting of low illumination with low CCT (Warm-Dark), low illumination with high CCT (Cold-Dark), high illumination with low CCT (Warm-Bright), and high illumination with high CCT (Cold-Bright) (Figure 2). The study used a within-subjects design; each participant experienced all 17 lighting conditions, with the order of the conditions randomized for each participant. Scorpio and colleagues. Scorpio and colleagues [115] conducted a study demonstrating that UE4 can accurately replicate light distribution when appropriately calibrated. Based on this study, a comparison of luminance values was made between UE4 and DIALux Evo.

**Table 1.** The luminous flux for 6 luminaires and the corresponding illuminance of the working plane.

| Lighting Condition | Luminous Flux (lm) | Average Illuminance (lx) |
| --- | --- | --- |
| N1 | 300 | 45 |
| N2, WD, CD | 1000 | 150 |
| N3 | 1700 | 290 |
| NM, M1-to-M6 | 2400 | 420 |
| N5 | 3100 | 550 |
| N5, WB, CB | 3800 | 670 |
| N6 | 4500 | 780 |

The adjustment phase of the experiment involved importing and simulating the same number of luminaires with a light distribution profile in UE4. To control the luminous flux of the light source and CCT, the right and left-hand controls were assigned, respectively. Specifically, they could increase or decrease the luminous flux and color temperature in increments or decrements of 100 lumens and 100 Kelvins, respectively. However, accurately reading the available illumination level (lux) at a surface, such as a desk, using UE4 version 4.26 was not possible. Therefore, the brightness was measured using the strength of the lighting source in lumens [116], as explained and compared in the Scorpio and colleagues' study [115] specifically in a subsection known as method three.

There were several reasons for choosing UE4 over other engines. Firstly, UE4 incorporates lighting algorithms based on physically based shading, which ensures accurate simulation of the interaction between light rays and surfaces, following the inverse square law [117] and including interactions between light and materials [118], [119]. Secondly, UE4 utilizes physically based lighting units [116]. Additionally, there was no need for additional simulation software or HDR cameras to assess or replicate the distribution of brightness within the virtual setting. This allowed individuals to enter the illuminated virtual environment and evaluate the lighting design from different perspectives without requiring extra equipment. This comprehensive evaluation enabled the assessment of user satisfaction, comfort, and interaction within the virtual environment. It is worth noting that simulating lighting in VR engines is a challenging task and an area of rapid development [120].

To enhance visual quality, UE4 version 4.26 employed pre-computed lighting and post-processing algorithms. Following the guidelines in the engine's documentation, the default luminance range was extended to establish the brightness conditions, enabling the use of correct lux values for lights in the scene while respecting auto exposure without causing the image to be overexposed [116], [121], [122].

For compatibility with VR head-mounted displays (HMD), a Tone Mapping Operator (TMO) was employed to convert High Dynamic Range (HDR) scenes to Low Dynamic Range (LDR). The TMO used in this experiment was similar to the one utilized in Hegazy and colleagues' study [122]. Although ongoing research explores various tone mapping techniques in lighting research using Immersive Virtual Environments (IVEs) ongoing [55], [61], [123]–[126], the Academy Color Encoding System (ACES) Filmic Tone Mapping Curve was employed, as it is the current default tone mapping curve in UE4 [127], [128]. The ACES Filmic tone mapping algorithms ensure consistent color preservation across different formats and display devices. This approach maintains color accuracy and future-proofs the source material, eliminating the need for constant adjustments for emerging mediums and saving time and effort [129]. In the developed system, the Filmic TMO was defined by the following parameters: Slope = 0.88, Toe = 0.55, Shoulder = 0.26, Black clip = 0.0, White clip = 0.04. Gamma correction (2.2) was implemented as part of the rendering pipeline in UE4 [130]. Considering the color space for scene projection on the HMD we used Quest Link. Therefore, it should be noted that the displays of Meta Quest 2 closely adhere to the RGB primaries of the Rec.709 color space while also employing a white point that closely approximates D75 [130].

While efforts were made to create realistic lighting simulations using UE4, it is important to acknowledge that further research, such as this study, is necessary to validate and discuss these findings in relation to real-world environments.

**Figure 2.** The 17 lighting conditions in the experiment. In the labeling schema, the CCT ranged from W=warm to N=neutral to C=cool, and the illumination ranged from D=dark to M=medium to B=bright.

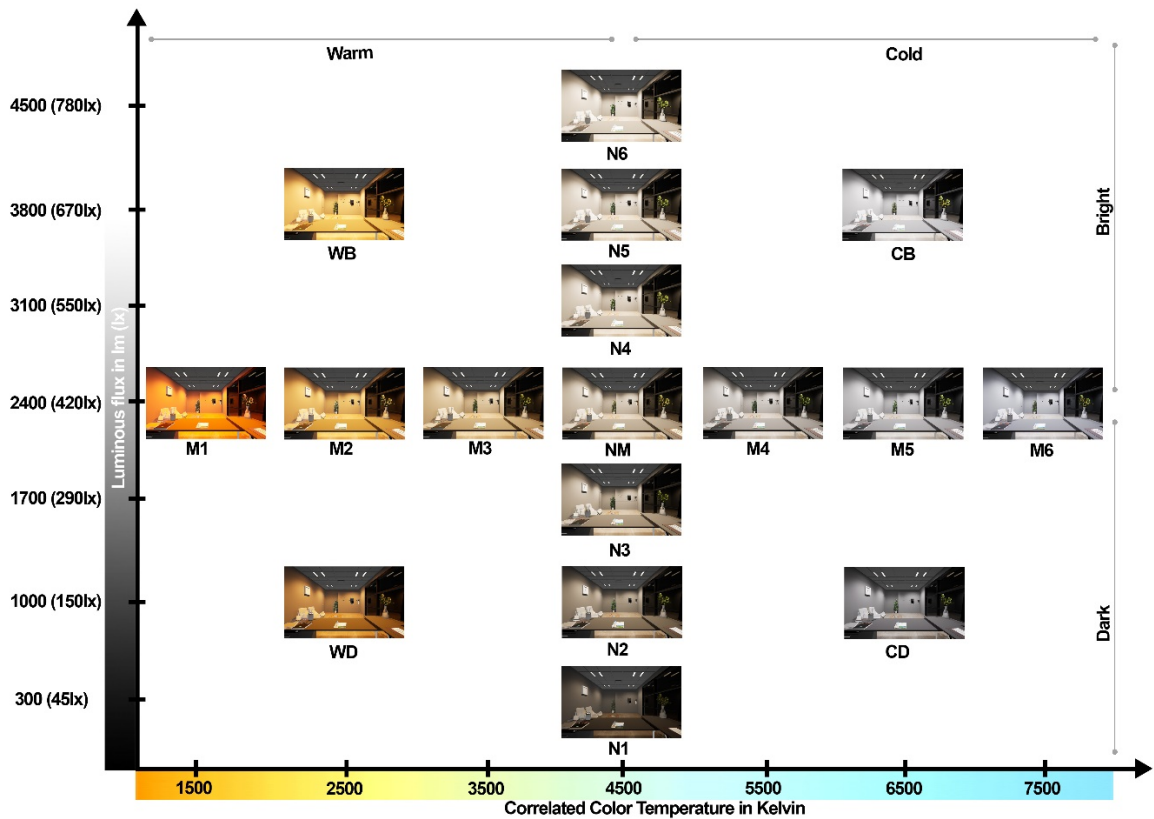

## 2.3 Participants and Sample Size

We've conducted a power analysis before recruitment. To examine the interaction effects (comparison of slopes between genders) and achieved a power of 0.80 at 0.05 alpha level assuming medium effect size (d=0.50), the required sample size would be 128. With 17 tasks per participant and assuming an Intra-Cluster Correlation (ICC) of 0.10, it would round up to 20 participants (effective sample size = 130.77), 10 per gender.

Thirty-three healthy adult participants were recruited using convenience methods (word-of-mouth and announcements on departmental e-mail lists). Data from 6 of the participants had to be excluded from the study due to technical problems (missing or incomplete data). Thus, the final sample included 27 participants. All participants were associated with **[deleted for the purpose of blind review]**. Twenty-three participants were undergraduate university students, with the remaining population being graduate students and faculty members. Sixteen (16) of the participants reported as female, and eleven (11) as male. The participants ranged in age from 18–64 years, with a mean of 24.16 years (SD 8.09). All participants affirmed that they had no medical problems with their eyes (cataracts, glaucoma, injury, etc.) and no difficulty in distinguishing colors. All of the participants reported getting a sufficient amount of sleep (7–9 hours) on the night before the experiment. The majority were using eyeglasses (59%), in which 75% of them had difficulty seeing far (such as road signs), and 18% had difficulty seeing far and close (such as reading a book). Most of the participants were students or researchers (66.6%) and programmers (25%), while a broad range of working tasks was represented, but the majority stated typing (92%), using a computer (77%), and reading (62%) as their daily job. Each participant gave informed written consent prior to the research activities, and the overall study protocol was approved by the Institutional Review Board (IRB) at **[deleted for the purpose of blind review]**.

**2.4 Measurement Tools and Metrics**

A diverse set of measurement tools was obtained from our previous studies. Some of these tools were questionnaires, while others required the researchers to complete them through observation.

*Questionnaires:* In this study, Demographic information was collected before the beginning of the experiment. The *Self-Assessment Manikin* (SAM) [131] was employed as a graphical assessment method to measure levels of arousal and emotional valence ranging from 1 (low) to 9 (high). Specifically, participants were instructed to rate their arousal level (ranging from calm to excited) by answering the question, "How alert does this lighting make you feel?" Additionally, they were asked to rate the emotional valence (ranging from happy to unhappy) by responding to the question, "Would you want to work in this lighting condition?" These measures were used to capture participants' subjective evaluations of their alertness and emotional response to the lighting conditions. We also used the *Morningness–Eveningness Questionnaire (MEQ)* [132], the *Igroup Presence Questionnaire (IPQ)* [133], the *Simulator Sickness Questionnaire (SSQ)* [134], and the *NASA Task Load Index* [135]. The level of fatigue was also measured (1-9 scale) both at the beginning and at the end of the experiment. The results and interpretation for these metrics other than SAM are reported in section 6.

*Lighting Adjustment Metric:* Aside from the final values obtained from adjustment tasks, we developed additional metrics, known as lighting adjustment uncertainty level, based on user lighting adjustment behavior which gives us further information about the quality of adjustment behavior using keystroke dynamics. The Adjustment Uncertainty (AU) refers to the total number of wrong clicks (button press) which are against the direction of the preferred LC divided by the total number of correct clicks which are in favor of the preferred LC. For instance, in an initial lighting condition of WD (1000 lm and 2500 K), if the final preferred LC for the participant is recorded at 1800lm and 5000 K (an upward trend for lumen and CCT), then the number of clicks decreasing the light intensity was divided by the number of clicks increasing the lighting intensity. If we consider the brightness, 8 clicks upward were needed to reach from 1000lm to 1800lm while the participant performed 10 upward and 2 downward clicks, which resulted in an illuminance uncertainty level of 0.2. The value is measured on a scale of 0 to 1, while 0 shows no uncertainty at all and 1 shows maximum uncertainty (i.e., returning to the initial condition after evaluation). This was calculated for the light illuminance intensity and CCT separately and showed how much each individual was unsure about his/her own decision during the adjustment task.

## 2.5 Study Procedure

After providing informed consent, the participants' experimental session was held at the **[deleted for the purpose of blind review]** study facility. Upon arriving for the session, each participant filled out a demographic form as well as MEQ. After wearing the VR-HMD, the researchers next instructed the participant on how to use the hand-held controllers. Next, the participant was given a three-five-minute VR tutorial to become comfortable with the lighting adjustment tools and to practice responding to verbal questions from the researchers. It should be noted that the initial lighting condition was NM, but the participant started to manipulate this condition promptly with the controllers for learning purposes. The duration of the VR session per experiment was around 17 minutes, excluding the tutorial.

After initiating the main experiment, participants first rated their level of mental fatigue, and they went through a loop of seventeen different lighting conditions with a devoted 50s time for each. The sequence of each condition started with a 10s of observation of the given lighting condition, followed by two Self-Assessment Manikin (SAM) questions for emotional valence and arousal. Participants were asked to verbally rate their emotion and arousal on a 1-9 scale after seeing the related widget and overhearing the voice. Simultaneously, the researcher typed the participant's answer using a keyboard, and our scripting allowed us to integrate this in automatically in the final log file. Next, they heard a voice that instructed them that adjusting the light to their preference is allowed using their controller. They were allowed to adjust the light for 25s for each lighting condition, and a black screen for 5s followed this before the next condition. At the end of all 17 lighting conditions, the level of fatigue was measured. Finally, each participant was asked to fill out a series of post-questionnaires, which assessed the level of immersion, cybersickness, and task load index.

## 2.6 Data Analysis Model

We used the R language with libraries "lme4", "emmeans", "effectsize" for our data analysis. For RQ1 and 2, due to the nested nature of the data, we fitted a linear mixed model (LMM) with the fixed effects of gender, Initial CCT, their interaction effect, and order while controlling for the random effect of participants to predict preferred CCT; and a linear mixed model with the fixed effects of gender, initial illuminance, their interaction effect, and order, while controlling for the random effect of the participant to predict preferred illuminance. For RQ3, we fitted linear models with the fixed effects of gender, initial CCT, initial illuminance, plus gender by initial CCT and gender by initial illuminance interaction effects and order while controlling for the random effect of the participant to predict valence and arousal. We did not expect any gender-by-order interaction effect.

For all the models fitted, we performed f-tests with Satterthwaite's method to examine the effects of predictors, and calculated effect sizes, partial $\eta^2$ and $\omega^2$. We then reported the estimated coefficients at different gender levels if the interaction effects were found to be significant.

## 3. Results

### 3.1 The Effect of Initial LCs on Preferred LCs (RQ1)

Figure 3 depicts the participants' preferred luminous flux and CCT values for each lighting condition. The purple spots represent the specific lumen intensities (Fig 3.A) and CCTs (Fig 3.B) associated with each condition, while the green dots indicate the mean values chosen by participants for each condition. The blue and red dots also represent the adjusted mean values for males and females. The adjustments regarding light intensity adjustment reveal that across all lighting conditions, participants generally preferred average values ranging from 1000lm (~150lx) to 2000lm (~300lx), while for CCTs, the preferred range was between 4000K and 6000K.

**Figure 3.** The lighting adjusted values for the given lighting conditions. **A.** luminous flux and **B.** CCT. Note: The ROUT method removed 2% of the outliers for the presentation.

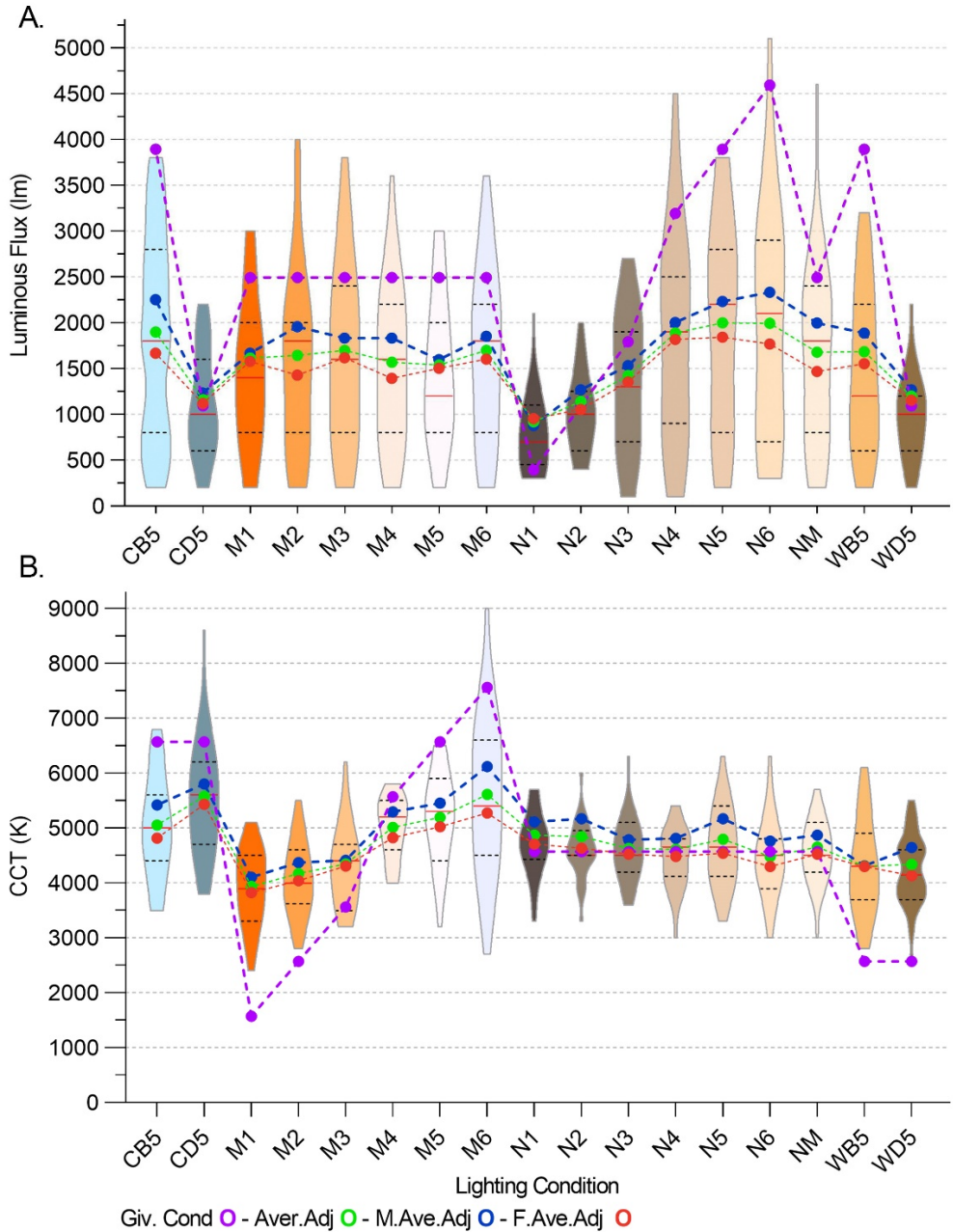

Table 3 shows the f-test results of the predictors for preferred lighting illuminance and CCT. We found the main effect of initial illuminance on preferred illuminance as well as initial CCT on preferred CCT significant. However, we did not find the main effects of gender on our dependent variables significant. Additionally, we found a significant gender by initial CCT interaction effect on the preferred CCT, $f(1, 429) = 6.71$, $p = .010$, partial $\eta^2 = .02$, $\omega^2 = .01$. For females, the coefficient of initial CCT on preferred CCT was estimated to be 0.24, SE=0.02, 95% CI: [0.20,

0.27]; for males, it was 0.31, SE=0.02, 95% CI: [0.27, 0.35]. The difference in coefficients between genders (Female – Male) was -0.07, SE = 0.03, 95% CI: [-0.13, -0.02].

Regarding the illuminance, we found a significant gender by initial illuminance interaction effect on the preferred illuminance, f(1, 429) = 9.23, p = .003, partial $\eta^2$=.02, $\omega^2$= .02. For females, the coefficient of initial illuminance on preferred illuminance was estimated to be 0.21, SE=0.03, 95% CI: [0.16, 0.26]; for males, it was 0.33, SE=0.03, 95% CI: [0.27, 0.39]. The difference in coefficients between genders (Female – Male) was -0.12, SE = 0.04, 95% CI: [-0.20, -0.04].

**Table 3.** F-Test results of the predictors of the preferred light illuminance and CCT.

| Predictor | Numerator DF | Denominator DF | F Value | P Value | Partial $\eta2$ | $\omega^2$ |
|---|---|---|---|---|---|---|
| **ANOVA Light CCT** | | | | | | |
| Gender | 1 | 35.196 | 0.035 | 0.852 | 0.001 | 0.000 |
| Initial CCT | 1 | 429.000 | 375.188 | 0.000 | 0.467 | 0.465 |
| Order | 1 | 429.000 | 2.959 | 0.086 | 0.007 | 0.005 |
| Gender: Initial CCT | 1 | 429.000 | 6.707 | 0.010 | 0.015 | 0.013 |
| **ANOVA Light Intensity** | | | | | | |
| Gender | 1 | 29.63 | 0.004 | 0.952 | 0.000 | 0.000 |
| Initial Intensity | 1 | 429.00 | 178.697 | 0.000 | 0.294 | 0.292 |
| Order | 1 | 429.00 | 19.081 | 0.000 | 0.043 | 0.040 |
| Gender: Initial Intensity | 1 | 429.00 | 9.232 | 0.003 | 0.021 | 0.019 |

**3.2 The Order Effects of ILC on Preferred Illuminance and CCT (RQ2)**

Figure 4 shows the preferred values for lighting illuminance and CCT with the increase of presented lighting order from 1 to 17. During the experiment, both males and females preferred significantly lower values for the lighting intensity and marginally significantly lower CCT as task order increased (table 3). The order of presented lighting condition was a marginally significant predictor of preferred CCT with an estimated coefficient of -7.82, SE=4.54, 95% CI: [-16.71, 1.08], f(1, 429)=2.96, p=.086, partial $\eta^2$<.01, $\omega^2$<.01. It was a significant predictor of preferred illuminance with an estimated coefficient of -19.69, SE=4.51, 95% CI: [-28.52, -10.87], f(1, 429)=19.08, p<.001, partial $\eta^2$=.04, $\omega^2$= .04.

**Figure 4.** Illustration of preferred lighting conditions by order. **A.** Illuminance and **B.** CCT

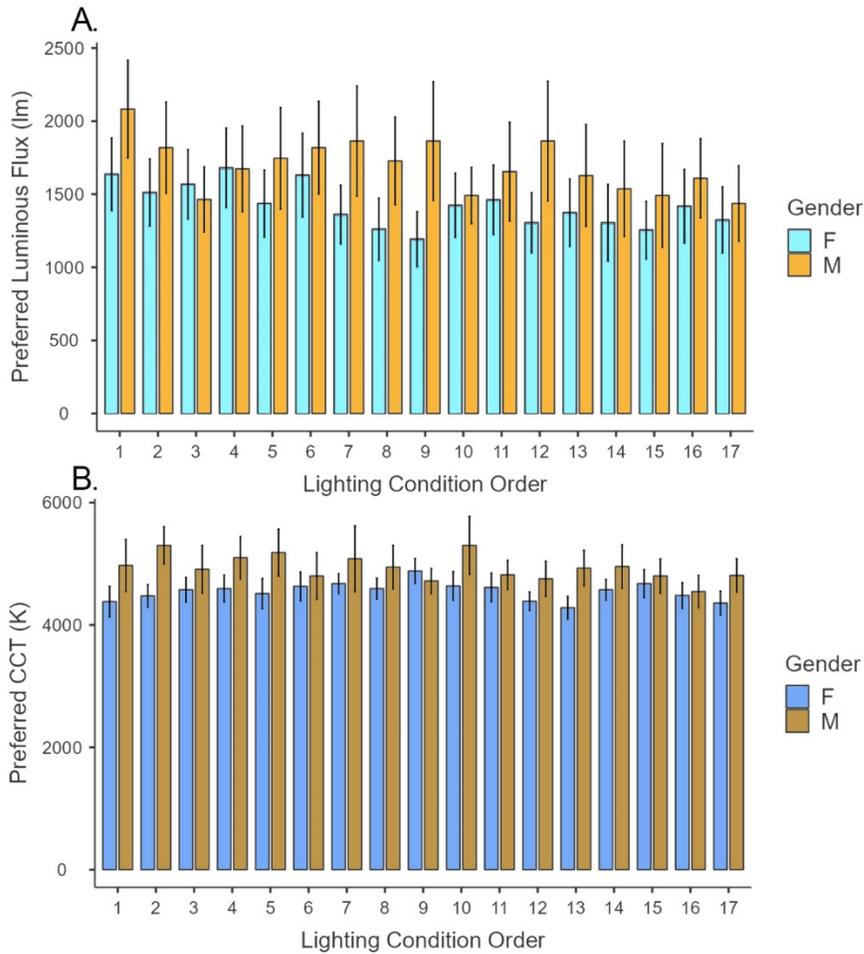

## 3.3 The effect of Initial Lighting Condition on Emotional Valence and Arousal moderated by gender (RQ3)

Figure 4 shows the valence-arousal diagram of lighting conditions used in this study. At 4500K N1 and N2 were found to be more relaxing and calmer while bright conditions (over 670lx) such as WB, CB, and N6 were rated as more alerting and tense.

**Figure 4.** Emotional Valence and Arousal Results 2D Model based on Russel's circular graph.

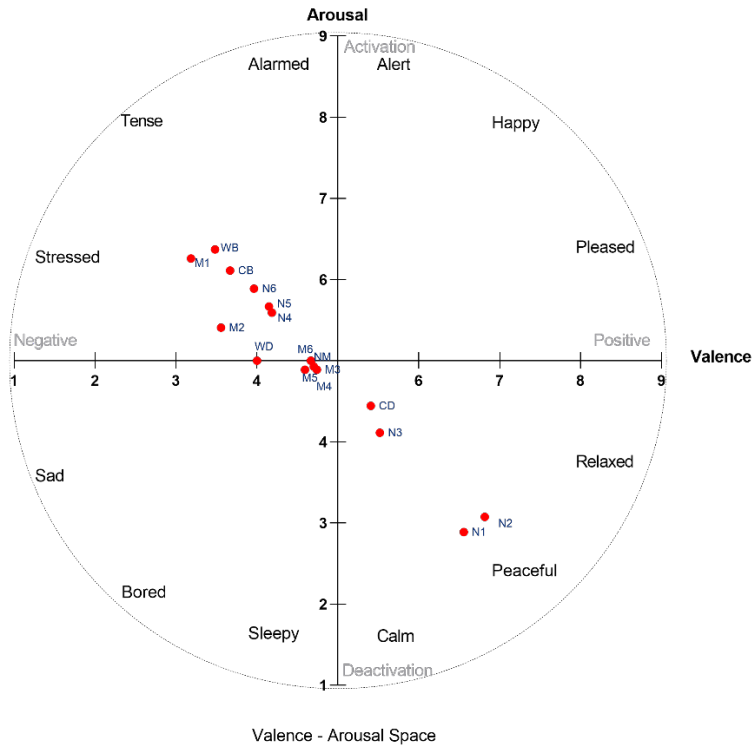

Figure 5 shows the impact of illuminance and color temperature of light on emotional valence and arousal levels across different genders. Regarding the effect of illuminance, females presented a decrease in the emotional valence mean score with an increase in illuminance level, moving from 7.00 at 45 lx to 3.19 at 780 lx. Similarly, males exhibited a decrease in valence from 5.91 at 45lx to 5.09 at 780lx (Fig 5.A). In terms of arousal, an increasing trend was observed for females with the increase in illuminance. The mean score rose from 2.50 at 45 lx to a peak of 6.63 at 3800 lx, falling slightly to 6.56 at 780 lx. Males presented a less consistent pattern, starting at 3.45 at 45 lx and ending at 4.91 at 780 lx (Fig 5.C).

Data on the CCT of light showed varying impacts on emotional valence and arousal for both genders. Females displayed an irregular pattern in emotional valence with changing color temperatures, reaching the highest mean score of 5.00 at 4500 Kelvin. On the other hand, males showed a slight increase in emotional valence from 2.82 at 1500 Kelvin to 5.45 at 7500 Kelvin

(Fig 5.B). As for arousal, females decreased from 6.13 at 1500 Kelvin to a minimum of 4.88 at 4500 Kelvin, then rebounded to 5.50 at 7500 Kelvin. Males showed a consistent decrease in arousal from 6.45 at 1500 Kelvin to 4.27 at 7500 Kelvin (Fig 5.D).

**Figure 5.** Level of emotional valence and arousal for the initial lighting illuminance and CCT for different gender

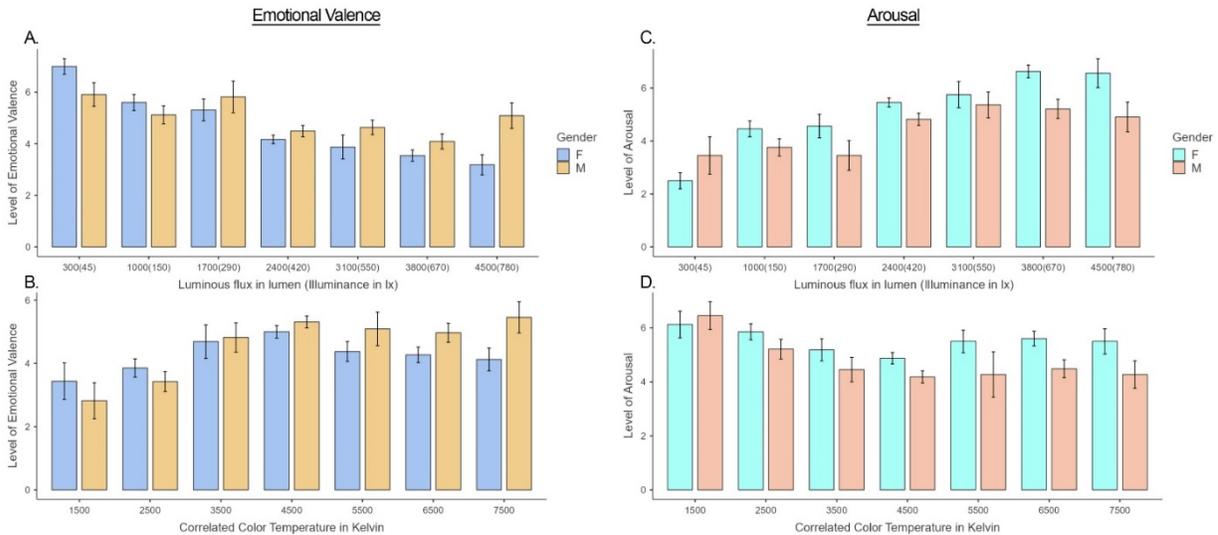

Table 4 shows the possible main and interaction effect for valence and arousal. For valence, we found a significant gender by initial CCT interaction effect, $f(1, 427) = 10.07$, $p = .002$, partial $\eta^2 = .02$, $\omega^2 = .02$, and a significant gender by initial illuminance interaction effect, $f(1, 429) = 13.76$, $p < .001$, partial $\eta^2 = .03$, $\omega^2 = .03$. The effect of order was not significant, $f(1, 427) = 0.03$, $p = .867$, partial $\eta^2 < .01$, $\omega^2 < .01$.

For arousal, we found a marginally significant gender by initial CCT interaction effect, $f(1, 427) = 3.86$, $p = .050$, partial $\eta^2 < .01$, $\omega^2 < .01$, and a significant gender by initial illuminance interaction effect, $f(1, 429) = 7.66$, $p = .006$, partial $\eta^2 = .02$, $\omega^2 = .02$. The effect of order was not significant, $f(1, 427) = 1.158$, $p = .282$, partial $\eta^2 < .01$, $\omega^2 < .01$.

Model estimated slopes by gender can be found in table 5.

**Table 4.** F-Test results of the predictors of emotional valence and arousal

| Predictor | Numerator DF | Denominator DF | F Value | P Value | Partial $\eta^2$ | $\omega^2$ |
|---|---|---|---|---|---|---|
| **ANOVA Emotional Valence** | | | | | | |
| **Gender** | 1 | 233.148 | 13.202 | 0.000 | 0.054 | 0.049 |
| **Given CCT** | 1 | 427.000 | 27.090 | 0.000 | 0.060 | 0.057 |
| **Given Intensity** | 1 | 427.000 | 70.993 | 0.000 | 0.143 | 0.140 |
| **Order** | 1 | 427.000 | 0.028 | 0.867 | 0.000 | 0.000 |
| **Gender:Given CCT** | 1 | 427.000 | 10.067 | 0.002 | 0.023 | 0.021 |
| **Gender:Given Intensity** | 1 | 427.000 | 13.765 | 0.000 | 0.031 | 0.029 |
| **ANOVA Arousal** | | | | | | |
| **Gender** | 1 | 128.002 | 1.854 | 0.176 | 0.014 | 0.007 |
| **Given CCT** | 1 | 427.000 | 12.267 | 0.001 | 0.028 | 0.026 |
| **Given Intensity** | 1 | 427.000 | 102.560 | 0.000 | 0.194 | 0.191 |
| **Order** | 1 | 427.000 | 1.159 | 0.282 | 0.003 | 0.000 |
| **Gender:Given CCT** | 1 | 427.000 | 3.861 | 0.050 | 0.009 | 0.007 |
| **Gender:Given Intensity** | 1 | 427.000 | 7.661 | 0.006 | 0.018 | 0.015 |

**Table 5.** Model estimated effects of the predictors of emotional valence and arousal by gender

| Predictor | Female | | | Male | | |
|---|---|---|---|---|---|---|
| | Coefficient | SE | 95% CI | Coefficient | SE | 95% CI |
| **Emotional Valence** | | | | | | |
| **Given CCT** | 0.096 | 0.06 | [-0.022, 0.215] | 0.396 | 0.073 | [0.253, 0.54] |
| **Given Intensity** | -0.820 | 0.086 | [-0.989, -0.650] | -0.319 | 0.104 | [-0.523, -0.114] |
| **Order** | -0.002 | 0.015 | [-0.032, 0.027] | -0.002 | 0.015 | [-0.032, 0.027] |
| **Arousal** | | | | | | |
| **Given CCT** | -0.072 | 0.059 | [-0.188, 0.045] | -0.254 | 0.072 | [-0.395, -0.113] |
| **Given Intensity** | 0.856 | 0.085 | [0.689, 1.022] | 0.488 | 0.102 | [0.287, 0.689] |
| **Order** | -0.016 | 0.015 | [-0.045, 0.013] | -0.016 | 0.015 | [-0.045, 0.013] |

## 3.4 User Lighting Preference Profile

Figure 6. illustrates the effects of initial illuminance and initial CCT on the level of uncertainty for preferred illuminance and CCT, respectively. Generally, both females and males experience a reduction in uncertainty as the lighting intensity increases. However, the pattern differs slightly between genders (Fig 6.A). For females, the uncertainty level begins at 0.6 (45 lx), and decreases to a minimum of 0.2 (780 lx). For males, the uncertainty level starts at 0.5 (45 lx), and then decreases to a low of 0.2 (780 lx).

Regarding the initial CCT, both females and males experience peaks in their uncertainty levels, followed by a decline as the CCT increases. Similarly, the pattern and the CCT at which the peak occurs differ between genders (Fig 6.B). For females, start with an uncertainty level of 0.3 (1500 K), the uncertainty rises to its peak at 0.6 (4500 K), and then gradually declines to 0.3 (7500 K). for males presents a slightly different trend. They begin with an uncertainty level of 0.2 (1500 K), which then rises to a peak at 0.5 (5500 K), and subsequently decreases to 0.3 (7500 K).

**Figure 6.** Level of calculated uncertainty across different **A.** ILC and **B.** CCT

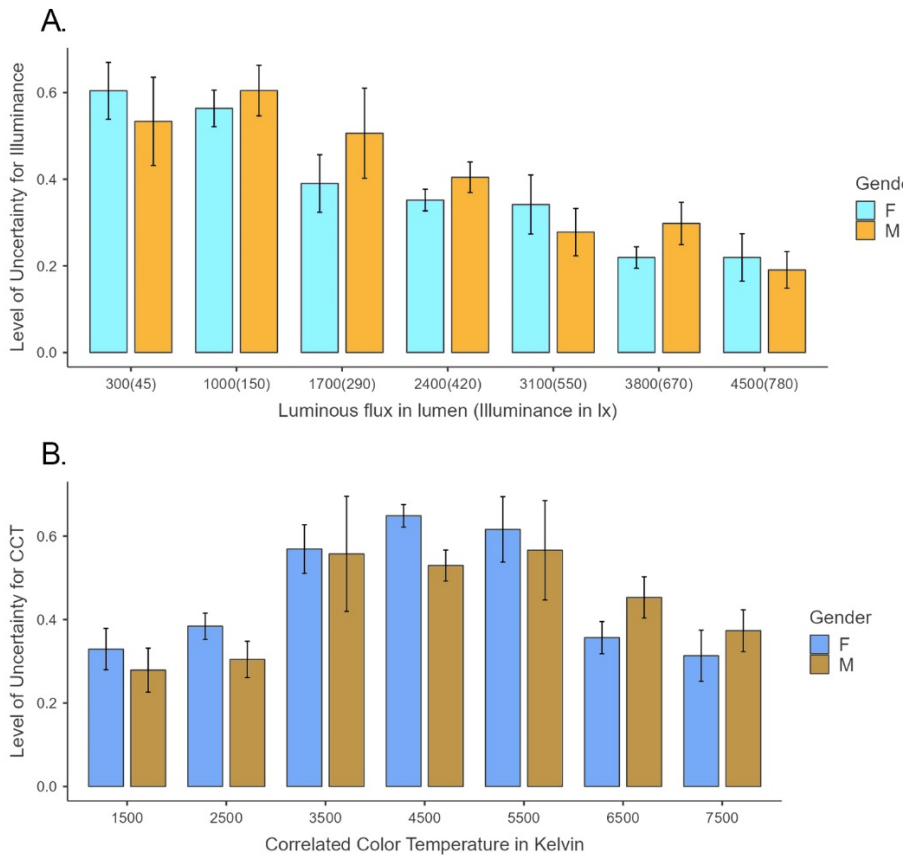

## 4. Discussion

In this study, we aimed to address several research questions. First, we focused on understanding the impact of given lighting brightness and CCT on preferred lighting through a VR adjustment task. Then we aimed to examine the influence of the order effect of the presented lighting condition on participants' preferences. Additionally, we explored whether adjusted lighting conditions were influenced by gender. Meanwhile, we also measured the participant's emotional valence and arousal during exposure to different lighting conditions. Lastly, we analyzed the user adjustment profile during the adjustment task.

### 4.1. Contribution to lighting research

Regarding the adjusted illuminance found in this study, the experimental results in VR align with those found in the real environment in Logadottir and colleagues' study [70]. The experimental findings confirmed that both the stimulus range and the anchor point employed in the adjustment task substantially impacted the participants' preferred illuminances. This highlights the importance of considering and reporting these variables when utilizing the adjustment task method to determine user preferences. The results further demonstrated that higher given illuminance levels led to higher preferred illuminance values in the adjustment task. Moreover, these results are in line with those found by Uttley and colleagues and Fotios and colleagues [101], [108] Therefore, our results from the adjustment task in VR also confirm that participants preferred significantly lower illuminances when presented with a lower range of lighting conditions. The findings in our study can be interpreted as similar to those found in de Kort and colleagues' study that [77] offering higher pre-set values results in higher preferred values. Regarding the CCT, our results found in this study align with those found in Logadottir and colleagues' study [65], indicating that the stimulus range used in the adjustment task significantly influences the preferred CCT chosen by participants. In other words, the range of CCT values presented to participants during the task affects their subsequent preferences for CCT. However, the average preferred CCT adjusted by all participants in VR (4700K) was slightly higher than those found in Viitanen and colleagues' experiment, which was 4150K [136].

Regarding the moderation effect of gender for preferred illuminance and CCTs, we found that the main effect of gender is not significant for illuminance nor CCT. However, we found significant interaction effects of gender by initial illuminance and CCT on the preferred illuminance and CCTs, respectively. These outcomes indicate that preferred values for each lighting condition differed between males and females. Notably, our result was consistent among almost all of the 17 lighting conditions in which men preferred a higher level of illuminance and CCT compared to women. In this regard, our results are similar to the finding of Schweitzer and colleagues' experiment [103] for the preferred values for activation and relaxation; males preferred higher illuminance levels in comparison with females. Regarding the lighting CCT, our results VR is in line with those found in Huang and colleagues' study [81]], in which women preferred lower CCT compared to men. Moreover, our results for both illuminance and CCT are similar to those found in Wang and colleagues' study [137] which shows that male observers preferred light sources with higher CCTs and illuminance, but female observers preferred light sources with relatively lower illuminance–CCT combinations. However, our results for the gender are in contrast with those findings in Veitch [138], Dangol and colleagues [139], Islam and colleagues [140], and Veitch and colleagues' experiments [29] which no statistically significant interactions between gender and light level and CCT were found.

While controlling the order effect in our LMM model (RQ2), we found that the main effect of the order on preferred light intensity was statistically significant, while this was marginally significant for the preferred CCT. The results also show that with randomized order for the presented lighting conditions, participants preferred lower light intensity for the later presented condition in general compared to the early ones. However, the order effect changes for preferred CCT were not as significant as illuminance. Our results in VR revealed substantial bias due to order effects for the preferred lighting conditions values. Results in this study are similar to those found in Kent and colleagues' studies [105], [107] for the randomized order of reported discomfort due to the glare. Even though we didn't incorporate glare discomfort in our study due to the limitation of VR, our results demonstrate the need for caution when interpreting subjective evaluations for the preferred CCT and illuminance level. The lighting adaptation, habituation, and fatigue in VR could explain the participant's preference for lower illuminance. Participants may have experienced adaptation to the previous lighting conditions, causing their visual system to become more sensitive to light. As a result, they may have perceived higher lighting intensities as excessively bright or uncomfortable, leading them to prefer lower values [108]. Participants might also have habituated to the lighting conditions over time. Initially, they may have been more cautious or conservative in their lighting intensity preferences, but as they became more accustomed to the task and the range of lighting conditions, they may have adjusted their preferences towards lower values. Lastly, engaging in a prolonged lighting adjustment task can be mentally and visually fatiguing. As the task progresses, participants may become more fatigued, as reported in section 6, leading to a preference for lower lighting intensities perceived as less visually demanding or tiring.

Our results on arousal and emotional valence rating for the given lighting conditions (RQ3) revealed that arousal and valence are significantly affected by both light illuminance and CCT. Statistical significances were also found for the interaction of given CCT and given illuminance by gender. Regarding the emotional valence to measure the level of happiness about the given brightness and CCT, participants reported their preference to work in the environment with lower brightness (45-290lx) in comparison with higher brightness (420-780lx), while the reverse trend was observed for CCT. In this context, our results are in line with those found in Chen and colleagues' study [141] which participants felt more comfortable under 50 lx and 100 lx instead of 500 lx regardless of the time of the day. The results regarding the valence are also in line with those found in Lee and Lee's experiment [142] which a higher CCT (Blue light) was reported as a higher level of pleasure in comparison with a lower CCT (red-yellow light).

Regarding arousal, participants rated their level of arousal higher in higher brightness (i.e., 45-290lx vs. 420-780lx), while the higher CCT did not result in significantly higher arousal. Our arousal results align with those found in Cai and colleagues' study [143] showing the increment trend of arousal with higher illuminance. The effect of higher brightness on arousal and alertness has been proved not only with subjective emotional measurements such as SAM in this study but also through physiological measurements [144]–[147]. The result of both emotional valence and arousal are in line with those subjective evaluations in Rajae-Joordens study [148] that red light is perceived as less pleasant and more arousing than blue light, which implies that red light evokes stronger levels of arousal and may stimulate a heightened sense of alertness or excitement compared to blue light. Moreover, in accordance with expectations, saturated light (colors with high intensity and purity, such as those in reddish hue) is rated as more arousing than desaturated light (colors with lower intensity or mixed with white). This indicates that the vividness and richness of the light color play a significant role in the level of arousal experienced. Overall, these findings emphasize the impact of both color hue (red vs. blue) and saturation on individuals' subjective experiences of pleasure and arousal in relation to different lighting conditions.

In this context, to discuss the gender effect moderation on arousal and valence by given Illuminance and CCT, the study reveals that both illuminance and color temperature significantly impact emotional valence and arousal. While both genders showed a decrease in emotional valence with increased illuminance, females displayed a greater change in arousal levels compared to males. As for color temperature, emotional valence for females showed no clear trend, while males showed a mild increase with higher color temperature. Arousal levels for both genders seemed to decrease with increasing color temperature, although the pattern was more consistent for males than females. These findings indicate the need to explore further the complex relationships between gender, environmental lighting conditions, and emotional responses. Given the differences in responses between genders, lighting design considerations for various environments should consider these gender-specific impacts.

Lastly, regarding the exploratory part of this study on the introduced metric for the level of uncertainty, for both illuminance and CCT, there is a pattern of uncertainty peaking at mid-range levels, followed by a decrease as the lighting parameter increases. However, the specific pattern and the level at which the peak uncertainty occurs differ by gender and whether the lighting parameter is illuminance or CCT. Females seem to have a higher level of uncertainty at lower illuminance levels and mid-range CCT values. This suggests that females may find adjusting lighting conditions at lower illuminance levels and in the mid-range CCT more challenging. Conversely, males face their peak uncertainty at mid-illuminance levels and mid-to-high CCT values, indicating that they may find it more challenging to make decisions in these specific ranges. These findings offer valuable insights for lighting design, particularly in contexts requiring user-controlled adjustments. Designers can use this information to create more intuitive interfaces and lighting systems that consider the variations in user response to different lighting conditions and CCT values.

**4.4. Implications for Practice**

In practical terms, our results indicate that lighting settings have profound implications for creating conducive environments for work or healing. Understanding user lighting preferences, especially in terms of illuminance and CCT, can provide a basis for more personalized lighting solutions, enhancing comfort and satisfaction. Moreover, the influence of gender on lighting preferences, as revealed in our study, underscores the necessity for more inclusive and gender-sensitive design considerations in lighting solutions.

The findings about the role of illuminance and CCT on arousal and emotional valence also have important applications. For instance, understanding how different light settings affect patients' emotional states could inform the design of healing spaces that foster relaxation and comfort in a hospital setting [97]. Similarly, in office environments, being aware of how specific lighting conditions influence arousal could aid in creating environments that promote productivity, focus, and general wellbeing.

The findings from this study emphasize the need for more in-depth and extensive research in this area. Future studies could focus on diversifying the participant sample, incorporating more real-world variables, and exploring the impact of natural light on preferred lighting conditions. This line of inquiry aims to enhance our understanding of how lighting design can optimize human comfort and wellbeing in various environments, be they offices, hospitals, or homes.

**4.3. Limitations**

The use of VR is a powerful tool for lighting research, as was mentioned in the Introduction, allowing high isolation and adjustment of variables that would be difficult or impossible to carry out in real-world environments. However, VR has some drawbacks because participant responses to these environments might not accurately reflect situations in real settings. Care should be taken when extrapolating the study's findings to non-virtual environments, especially in light of the fact that current VR systems do not provide great dynamic range compared to the human eye.

The current status of computer-generated environments utilizes a variety of rapidly evolving lighting simulation methods. We have encountered a couple of updates by UE4 developers for lighting simulation improvement during the experiment design, which continues to this day. Even though this sounds promising for the use of VR in lighting studies, it will create a barrier to generalizing our results or comparing them with those created in other game engines, such as Unity 3D or created with High Dynamic Range Images (HDRI) with different tone mapping algorithm [58], [61]. Although real vs. VR lighting comparison was not in our scope, designing and adjusting the task with all possible lighting conditions using HDRI is challenging. It's important to note, however, that while our study contributes to a growing body of research, it also emphasizes the complexities of the relationship between lighting conditions and human responses. Careful consideration of these dynamics is essential when translating our results into practical applications. Furthermore, while the use of VR in our study allowed for a high level of control over the variables tested, there are limitations to its real-world applicability. The nature of VR, the limitations of dynamic range in current VR systems, and the rapid evolution of lighting simulations should all be taken into account when translating the findings to actual environmental applications.

The loss of data from multiple participants, which led to a higher-than-expected attrition rate, was a notable problem we ran into during the study. Participants were excluded primarily because their VR log file data was corrupted or useless because timestamps were not printed, which were essential for data analysis. During the experiment, we encountered some technical difficulties with HMD battery crashes. The majority of these issues were ultimately caused by researcher error, which we will try to minimize in future research.

Last, but not least, it is important to point out that the study employed a convenience sample made up primarily of undergraduate students at **[deleted for the purpose of blind review]** and that we did not assess demographic variables or other elements that may affect the outcomes

(such as participant sleep schedules or time of day). These elements, particularly in light of participant diversity or similarity, may impact the preferred illuminance and CCT results.

### 4.4. Future Work

In an extension of this research trajectory, we will look at data for subjects who are involved in various kinds of activities under various lighting conditions. This will make it possible for us to thoroughly assess how well lighting and user preferences relate to certain activities, like memory, creativity, socializing, or relaxation in different architectural contexts. Additionally, as most of the interior spaces are lit by a combination of natural and artificial lighting, we can involve natural light in our future research. Future studies in this field should aim to evaluate larger participant samples and broaden the diversity of the participants in order to increase the generalizability of the findings and the possibility of examining fine-grained demographic and personal factors in relation to lighting preferences and responses.

Another way to measure preferred lighting conditions is through the use of biometrics. Biometric data, such as pupillary response, heart rate, and skin conductance, can provide insight into how different lighting conditions affect users physiologically. For example, a high heart rate or increased skin conductance may indicate that a particular lighting condition is causing stress or discomfort. Lastly, the role of natural light can't be ignored in predicting preferred lighting conditions. By taking into account the natural light available in a space and adjusting artificial lighting accordingly, researchers can create comfortable and preferred lighting conditions.

### 5. Conclusions

In conclusion, our study has made significant strides in understanding how various factors, such as illuminance, color temperature, order of presentation, and gender, can affect the preferred lighting conditions of individuals in virtual reality environments. By closely analyzing user adjustment profiles, we have gathered valuable insights that can help to optimize lighting design in various contexts.

One of the key findings is the substantial impact of the stimulus range and anchor point on participants' preferred illuminance and color temperature. Higher pre-set values of illuminance and color temperature led to higher preferred values, shedding light on how people adjust the lighting in VR settings. Our results also illustrated a gender discrepancy, showing that men prefer higher levels of illuminance and CCT than women. Additionally, the randomized order of lighting conditions presented influenced participant preferences, with lower light intensities generally favored for later-presented conditions. Significant influences of light illuminance and color temperature were also observed on emotional arousal and valence. The gender-specific impacts in these areas call for a more inclusive and gender-sensitive approach to environmental design. Furthermore, the variation in user response and level of uncertainty regarding different lighting conditions revealed through this study provide designers with valuable insights to create more intuitive interfaces and lighting systems.

## 6. Appendices

### 6.1 Descriptive Statistics and Potential Confounding Variables

The average score on *Mornignness-Eveningness Questionnaire* was 48.66 (SD=9.06), with the majority of intermediate (66%), a minority of moderate evening (22%), and moderate morning (11%). An intermediate score on the MEQ indicates a moderate preference for neither morningness nor eveningness, suggesting a flexible and adaptable chronotype. The *igroup Presence Questionnaire (*1-7 scale), the total score was 4.1 (SD=0.95), while the mean scores were 5.18 (SD= 1.17) for the sense of presence in VE, 4.72 (SD= 0. 6) for spatial presence, 3.8(SD=0.35) for Involvement, and 3.5 (SD=1.34) for experienced realism. These above-average results may have positive implications for the effectiveness of our VR environment. The average total score for the *Simulator Sickness Questionnaire* was 29.48 [out of 235.62], showing little amount of cybersickness reported in VR. This includes 15.19 for nausea, 32.12 for oculomotor disturbance, and 28.57 for disorientation. The results for the NASA Task Load Index (1–20 scale) indicated a low average score of 4.62 (SD=0.98). The level of fatigue (from 18 participants) was 4.29 and 4.72 out of 9 before and after the experiment, respectively.

**Acknowledgments:** The authors thank the team at the Design and Augmented Intelligence Lab at Cornell University, including, Jesus G. Cruz-Garza, Qi Yang Angella Lee, Elita Gao, Calvin Qui, , Talia Fishman, for assisting in data collection, recruitment, and brainstorming session.